\documentclass[12pt]{iopart}

\usepackage{amssymb}
\usepackage{stackrel}
	\usepackage{makeidx}
	\usepackage{amsfonts}
	\usepackage[ansinew]{inputenc}
	\usepackage[usenames,dvipsnames]{pstricks}
	\usepackage{subfigure}
	\usepackage{epsfig}
	\usepackage{pst-grad} % For gradients
\def\ba{\begin{eqnarray}}
\def\ea{\end{eqnarray}}

\def\bp{\mathbf{p}}
\def\br{\mathbf{r}}

\newcommand{\be}{\begin{equation}}
\newcommand{\ee}{\end{equation}}
\def\ba{\begin{eqnarray}}
\def\ea{\end{eqnarray}}
\def\bea{\begin{eqnarray}}
\def\eea{\end{eqnarray}}

\def\bp{\mathbf{p}}

\newcommand{\refer}[1]{(\ref{#1})}
\newcommand{\dd}{\mbox{d}}

\begin{document}
\title{Gravitational lensing of massive particles in Schwarzschild gravity}

\author{Xionghui Liu$^1$, Nan Yang$^2$, Junji Jia$^1$}

\address{$^1$ Center for Theoretical Physics \& School of Physics and Technology, Wuhan University, Wuhan, 430072, China}
\address{$^2$ Glyn O. Phillips Hydrocolloid Research Centre, Hubei University of Technology, Wuhan 430068, China}

\ead{junjijia@whu.edu.cn}

\begin{abstract}
Both massless light ray and objects with nonzero mass experience trajectory bending in a gravitational field. In this work the bending of trajectories of massive objects in a Schwarzschild spacetime and the corresponding gravitational lensing effects are studied. A {\it particle sphere} for Schwarzschild black hole (BH) is found with its radius a simple function of the particle velocity and proportional to the BH mass. A single master formula for both the massless and massive particle bending angle is found, in the form of an elliptic function depending only on the velocity and impact parameter. This bending angle is expanded in both large and small velocity limits and large and small impact parameter limits. The corresponding deflection angle for weak and strong gravitational lensing of massive particles are analyzed, and their corrections to the light ray deflection angles are obtained. The dependence of the deflection angles on the source angle and the particle speed is investigated. Finally we discuss the potential applications of the results in hypervelocity star observations and in determining mass/mass hierarchy of slow particles/objects.
\end{abstract}


\submitto{\CQG}

\maketitle

For a test particle with nonzero mass, it is known in both Newtonian gravity and general relativity (GR) that when it passes by a massive gravitational center (GC), its trajectory will bend. For massless particles, their trajectories will also bend because of GR. When particles from a fixed source are bent by the same center and reach an observer to form images (or even a ring), the effect of gravitational lensing (GL) happens.

The bending of light ray was first proposed by Einstein himself and served as one of the major evidence of GR in its early years \cite{einstein}. After the discovery of multiple images of quarsars \cite{Walsh:1979nx} and the first observation of giant blue luminous arcs \cite{lynds}, the weak GL phenomena was well established experimentally. More recently, even the Einstein ring \cite{hewitt} and weak GL of CMB \cite{Smith:2007rg, Das:2011ak, vanEngelen:2012va} and supernova lights \cite{Quimby:2013lfa, Nordin:2013cfa} have been observed. On the other hand, it is also found that weak GL has important observational consequences and theoretical applications. The GL was used to constrain cosmological parameters such as large scale structure etc (for a review see Ref. \cite{Lewis:2006fu}), as a probe of dark matter substructure \cite{Metcalf:2001ap, Metcalf:2001es}, coevolution of supermassive black holes and galaxies \cite{Peng:2006ew},  and to discriminate alternative gravitational theories.

When the particle rays pass by a very massive GC, the bending of the trajectory can be very large (exceeding $2\pi$ if the trajectory is close enough to the GC) and hence the strong GL might happen. Theoretically Darwin first studied the null geodesics in the strong field limit of Schwarzschild black hole \cite{darw1, darw2}. The strong GL equation of light ray then was studied by Virbhadra and Ellis \cite{Virbhadra:1999nm, Virbhadra:2002ju, Virbhadra:2008ws}. Frittelli, Kling and Newman established an exact lens equation and examined the Schwarzschild case \cite{Frittelli:1999yf}. Bozza et al investigated the same problem \cite{Bozza:2001xd} and then extended their work to allow more general spacetime \cite{Bozza:2002zj} and GL geometry \cite{Bozza:2009yw}. Perlick studied the exact GL with arbitrary bending angle \cite{Perlick:2003vg, Perlick:2010zh}.
Experimentally, even though no conclusive observation of strong GL of this kind  has been reported, it is still believed that this happens for example when particle rays pass by black holes located in the center of galaxies, or naked singularities and other exotic compact massive objects. The observation of strong GL will test not only the validity of GR but also its various alternative theories in the strong field limit.

Trajectory bending due to gravity not only happens to light or massless particles, but also to particles with mass. Examples of massive particles whose trajectories can go through a bending process range from hypervelocity stars (HVS) around massive black holes at the center of galaxies \cite{Yu:2003hj, Brown:2005ta}, to neutrinos and cosmic rays from stars and supernova, and to theorized WIMP and Axions (see Ref. \cite{Patla:2013vza} and references therein). Due to the strong gravity near the black holes and the large velocity of the HVS (up to $0.04c$), the trajectory bending of these stars obtained using GR will differ from the calculations made using only Newtonian gravity and therefore deserves a full GR treatment. For neutrinos, it is known that they are emitted along photons during the lifetime of stars and most vastly during supernova. The later together with active galactic nuclei (AGN) are also the sources for cosmic rays and ultrahigh energy neutrinos. Since the gravitationally lensed supernova has been detected \cite{Quimby:2013lfa, Nordin:2013cfa}, it is natural to expect that the neutrinos and cosmic rays emitted by these supernova and AGN are also lensed. Therefore studying the lensing of these particles will shed light on the properties of not only the particles involved, but also the lens and lensing mechanism themselves.

There are a few works that treat the GL of massive particles. Escribano et al. \cite{Escribano:2001ew}, Mena et al. \cite{Mena:2006ym}, Eiroa and Romero \cite{Eiroa:2008ks} have studied the GL of neutrinos by various objects, stressing on different aspects of the lensing effect but all using the bending formula of light rays. Even though the typical energy range of the standard model neutrinos considered by these authors might forces their velocity  to be very close to light, in principle the bending angles of massless and massive particles are different. Accioly and Ragusa studied the trajectory bending of massive particle with velocity close to the speed of light \cite{Accioly:2002ck} and found the bending angle to the second post-Newtonian order.
For heavier particles with even lower velocities, such as protons with low energy in the cosmic ray flux, their bending angle should be very different from that of light rays. Any application of the trajectory bending of these particles requires a more proper treatment of the bending angle. Very recently, the bending angle of massive particle in Schwarzschild metric was obtained by Tsupko \cite{Tsupko:2014wza} as a elliptic function of the closest distance and the energy or angular momentum of the particle. We emphasis that the bending angle in this work differ from Ref. \cite{Tsupko:2014wza} in that (1) we show the continuity of the bending angle from subluminal speed to the speed of light, and (2) our formula of the bending angle is given in terms of velocity and impact parameter of the particle, which is more convenient for the application in GL. Most importantly, the bending angle is only {part of} this work, while its applications in various GL scenarios constitute the major sections of the paper.

This work is organized as the following. In section \ref{cont} we will first show that in the Schwarzschild spacetime, particle trajectories as determined by the geodesic equations, are indeed continuous functions of the particle speed $v$ for $v$ from less than 1 to 1 (we set $G=c=1$ in this work). That is, the trajectory of massive particle can be continued smoothly to that of a massless particle when $v_{m\neq 0}/c$ approaches 1. We also show the existence of a particle sphere, an analogy to the photon sphere for the light ray bending and inside which particles incoming from far away will not loop out to reach an observer at infinity. In section \ref{gbfor} we then integrate the geodesic equations to produce an exact formula for the bending angle in the form of an incomplete elliptical function. Its fast and slow particle limits and weak and strong lensing limits are studied in sections \ref{wlur} and \ref{slur}.  In section \ref{vcor1}, these results are applied to the GL to show how the velocity correction can affect the deflection angles. Finally in section \ref{discus} we discuss the potential application of our results.

\section{Trajectory continuity from $v/c<1$ to $v/c=1$. \label{cont}}

We consider a bunch of particle rays coming from far away and eventually reach the observatories. During prorogation, they pass a static and spherically symmetric Schwarzschild gravitational field created by a GC and their trajectories are bent. The angular and radial geodesic equations that describe the trajectory take the form
\bea
&& \frac{\dd \phi}{\dd t}=\frac{L}{r^2}, \label{angt} \\
&& \frac{1}{2}\left(\frac{\dd r}{\dd t}\right)^2=\frac{1}{2} E^2-V(r), ~\mbox{where } V(r)=\frac{1}{2}\left(1-\frac{2M}{r}\right)\left(\frac{L^2}{r^2}+\kappa\right),\label{radt}\eea
and $\phi$, $r$ and $t$ are the metric coordinates. Here $M$ is the GC mass and $\kappa=1,~0$ for massive and massless particle respectively. $E$ and $L$ are the energy and angular momentum (per unit mass) of the particle at infinity. For massive particles, $E$ and $L$ can be expressed as
\be E=\frac{1}{\sqrt{1-v^2}},~ L=|\bp\times \br|=\frac{v}{\sqrt{1-v^2}}b, \label{eandlrel}\ee
where $b$ is the impact parameter and $v$ is the velocity of the particle at infinity. For both massive and massless particles, therefore the following relation holds $L/E=bv$.

Two facts are worthy noticing here. The first is that these geodesic equations form a system of first order differential equations. Physically it is expected that it permits a unique solution if and only if the initial position $(r_i,\phi_i)$ and initial velocity $(\dd r/\dd t|_i, \dd \phi/\dd t|_i)$ are fixed. In these initial parameters, because one can always add a constant to the azimuth angle, we can always fix $\phi_i$ as any constant we like. Fixing the initial velocity is equivalent to fix its initial speed $v$ and direction of velocity, which is characterized by the impact parameter $b$. Besides, the other two parameters in the geodesic equations, the energy $E$ and angular momentum $L$ per unit mass are also fixed by $v$ and $b$ as in Eq. \refer{eandlrel}.
Therefore it is clear that the trajectory will be completely determined by $v$, $b$, $r_i$ and the only parameter left in the geodesic equations, the GC mass $M$. The second point we note is that when $L^2>12M^2$, the potential term $V(r)$ in radial equation \refer{radt} has a maximum at
\be r_{-}=\frac{1}{2}\frac{\left(L-\sqrt{L^2-12M^2\kappa}\right)L}{M\kappa}, \ee
which equals the known result $r_-=3M$ for photons when $\kappa=0$.
Then to ensure that the trajectory is unbounded, i.e., an incoming particle will eventually come out, the $\dd r/\dd t$ has to flip sign from $-$ to $+$ at some point. From eq. \refer{radt}, this requires
\be \frac{1}{2}E^2<V(r_-),\label{cond1}\ee
a condition that will become important in the following sections.

We now show that if we do the following change of variables,
\be
r\to w=\frac{M}{r}, ~ b\to f\equiv \frac{bv}{M}, \label{wrbfrel}
\ee
the angle coordinate $\phi(w)$ for a massive particle traveled to position $w$ can be expressed as a function of three parameters: the initial $w_i\equiv M/r_i$, the speed $v$ and the ratio $f$. To do this, we square Eq. \refer{angt}, divide it by \refer{radt} and then change variables according to Eqs. \refer{eandlrel},  \refer{wrbfrel}. This establishes the following equation for $\phi$ in terms of $w$
\be \left(\frac{\dd\phi(w)}{\dd w}\right)^2=\frac{f^2}{[w^2f^2+\kappa(1-v^2)](2w-1)+1} .\label{mang}\ee
For $\kappa=1$, this equation can be readily solved to obtain
\be \phi(w)=\int_{w_i} \frac{f \dd w}{\sqrt{(w^2f^2-v^2)(2w-1)+2w}}. \label{mangsol}\ee
From this, the dependence of $\phi(w)$ for a nonzero mass particle on three variables $w_i$, $f$ and $v$ becomes very clear.

Next we show that $\phi(w)$ is continuous when the considered particle changes from an ultra-relativistic particle with $v$ approaching 1: $v/c<1$ to a massless particle with $v/c=1$. The corresponding angle function $\phi(w)$ for a massless particle can be obtained from Eq. \refer{mang} by substituting $\kappa=0$. The result is
\be \phi(w)=\int_{w_i} \frac{f\dd w}{\sqrt{w^2f^2(2w-1)+1}}. \label{m0ang} \ee
Comparing with Eq. \refer{mang}, one see that Eq. \refer{m0ang} coincides exactly with the smooth continuation of $v$ to 1 in Eq. \refer{mang}.

This continuity suggests that for ultra relativistic particles with nonzero mass, their bending angle is close to that of photons. Any measurable quantity that is related to the trajectory bending and GL for these ultra relativistic particles, such as deflection angle, magnification and time delay, will also be close to the corresponding values of light rays. The exact values of these variable should be approximatable by simple series expansions in $v$ at $v=1$, the leading order of which is just the result for light ray. In next section, we first derive a formula of bending angle for general $v$, and then present its leading order results.

\section{Bending of particles with general velocity $v$ \label{gbfor}}

The trajectory starting from a source $(w_i,~\phi_i)$ and ending at a detector at $(w_f\equiv {M}/{r_f},~\phi_f)$ will experience an bending angle of the amount $\Delta\phi=\phi_f-\phi_i$. In practice, usually the source and the detector are far away from the GC and therefore $w_i$ and $w_f$ are set to zero. In this case, using Eq. \refer{mangsol}, the bending angle becomes
\bea
\Delta\phi&=&2\int_0^{w_1}\frac{f\dd w}{\sqrt{(w^2f^2-v^2)(2w-1)+2w}} \label{manginw1}\\
&=&2\int_0^{w_1}\frac{f\dd w}{\sqrt{(w-w_1)(w-w_2)\left[2(w+w_1+w_2)-1\right]}}. \label{dpintform1}
\eea
The modulated bending angle, which measures the net change of direction of the ray, is defined as $\Delta\phi$ modulating away $(2n+1)\pi$
\be \Delta\phi_{\mbox{\scriptsize mod}}=\Delta\phi-(2n+1)\pi. \ee
Here $n$ are integers such that $|\Delta\phi_{\mbox{\scriptsize mod}}|\leq \pi$. $n>0$ corresponds to the case that the trajectory has looped around the GC. $w_1=M/r_1$ corresponds to the minimal radius $r_1$ of the trajectory, which can be solved as the smaller positive root of Eq. \refer{mang} when $\dd w/\dd \phi=0$. In terms of the new variables, this equation is exactly the denominator part of the integrand in Eq. \refer{manginw1}
\be (w^2f^2-v^2)(2w-1)+2w=0. \label{cubeq}\ee
$w_2$ in Eq. \refer{dpintform1} is the negative root of the above equation. Explicitly, $w_1$ and $w_2$ can be solved from Eq. \refer{cubeq} as functions of $v$ and $f$
\be w_{1,2}=\frac{\sqrt{f^2-12(1-v^2)}}{3f}\cos(\theta\pm\frac{2\pi}{3})+\frac{1}{6},\label{w1fv}
\ee
where
\be \hspace{-1.5cm} \theta=-\frac{1}{3}\arctan\frac{6\sqrt{3f^4v^2-(24v^4+60v^2-3)f^2 +48(v^2-1)^3}}{f(f^2-36v^2-18)}
\ee

\begin{figure}[htbp]
	\begin{center}
		\includegraphics[width=0.4\textwidth]{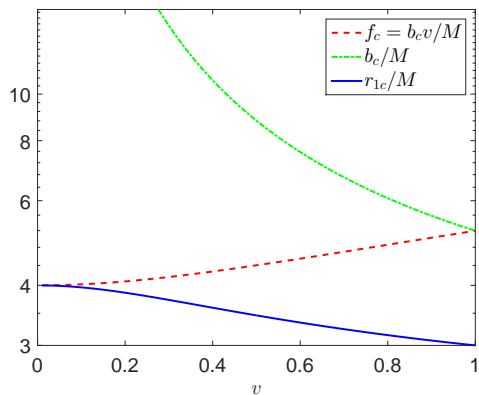} %depending on the latex compiler, you can omit the file extension
		\caption{The critical impact parameter $b_c$, the critical ratio $f_c$ and the radius of the particle sphere $r_c$ in terms of the GC mass. }
		\label{fcw1}
	\end{center}
\end{figure}

These roots are guaranteed to exist because of the condition \refer{cond1}, whose form in terms of $f$ and $v$ can be rewritten as
\be f\geq \frac{\left[8v^4+20v^2-1+(8v^2+1)^{3/2}\right]^{1/2}}{\sqrt{2}v}\equiv f_c \label{fcdef}.\ee
%\markred{
The $f_c$ here as a function of the velocity $v$, indeed is the critical ratio below which the particle will be captured by the GC. In turn, this defines through Eq. \refer{wrbfrel} a critical impact parameter
\be b_c\equiv \frac{f_cM}{v}=\frac{\left[8v^4+20v^2-1+(8v^2+1)^{3/2}\right]^{1/2}M}{\sqrt{2}v^2} \label{critip}\ee
and through Eq. \refer{w1fv} and Eq. \refer{wrbfrel}  a corresponding $r_{1c}$ at $f=f_c$
\be r_{1c}=\left(2+\frac{4}{\sqrt{8v^2+1}+1}\right)M. \label{r1cres}\ee
This formula was also obtained in Ref. \cite{Chandrasekhar:1985kt}. It defines a sphere that is analogous to the photon sphere in the case of light ray bending. We call it the {\it particle sphere}. Particles with velocity $v$ at infinity and with impact parameter smaller than $b_c$ will be captured eventually by the GC if it is only under the influence of gravity once it enters this sphere. Noticing the continuity of the geodesic rays from $v<1$ to $v=1$ as shown in the previous section and the existence of photon surface surrounding BHs as shown in Ref. \cite{Claudel:2000yi}, we believe that there should also exist particle surfaces for each subluminal velocity that surrounding all BHs. The critical ratio $f_c$ and particle sphere radius $r_{1c}$ is plotted in Fig. \ref{fcw1}. At the speed of light, one can easily check that formula \refer{critip} produces $b_c=3\sqrt{3}M$  and then Eq. \refer{r1cres} yields $r_{1c}=3M$, recovering the corresponding values for light ray. As the velocity decreases however, it is seen that the critical ratio $f_c$ is indeed reduced and the particle sphere radius is increased until they meet each other at $f_c=4=r_{1c}/M$ for $v\to 0$.

\vspace{5mm}

\begin{figure}[htbp]
	\begin{center}
\includegraphics[width=0.6\textwidth]{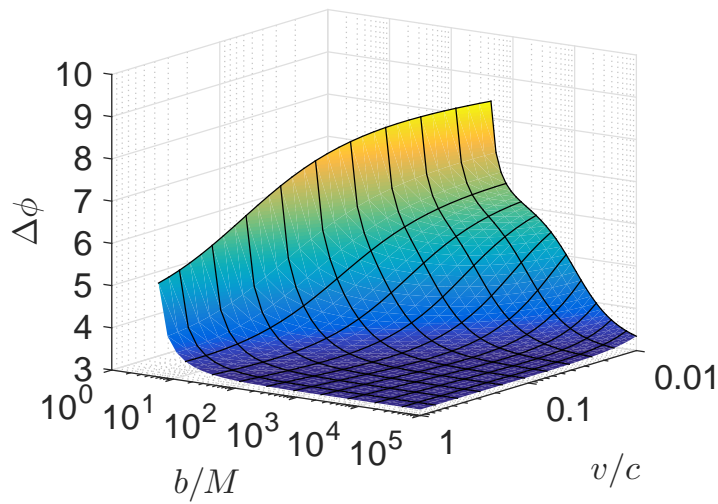}\\
	\caption{The bending angle as a function of $b_M$ and $v$. Note that at small $b$ or small $v$, the angle can exceed $2\pi$ and therefore loop around the GC. }
		\label{figdphi}
	\end{center}
\end{figure}

The angle $\Delta\phi$ in Eq. \refer{dpintform1} can be integrated out to obtain a function of only $f$ and $v$. However because the physical meaning of $f$ is not as simple and apparent as the impact parameter $b$ or $b/M$, hereafter we will mainly use the variable $b_M\equiv b/M$ when studying the bending angle and its various expansions. The critical impact parameter $b_c$ in Eq. \refer{critip} then will define a critical ratio $b_{Mc}\equiv b_c/M$. Using these variables, the bending angle \refer{dpintform1} can be worked out as
\bea
\Delta\phi(b_M,v)&=&-\frac{2\sqrt{2}i}{\sqrt{w_1-w_2}}F \left(\sqrt{\frac{2w_1}{h}},\sqrt{\frac{h}{2(w_1-w_2)}}\right), \label{exactgeneral}\eea
where $F$ is the incomplete elliptic function of the first kind, $h=4w_1+2w_2-1$ and $w_1$ and $w_2$ are functions of $b_M$ and $v$ given by Eq. \refer{w1fv} in which $f=b_Mv$ should be used. An essentially equivalent formula was obtained in Ref. \cite{Chandrasekhar:1985kt}. It is also given in Eq. (29) of Ref. \cite{Tsupko:2014wza}, which is given in terms of $r_1$ and $L$ (or $E$). The bending angle \refer{exactgeneral} applies to particles of any velocity and any impact parameter $b_M$ that is above its critical value.  If $\Delta\phi$ is close to $\pi$, then the bending is small and we say that the particles are weakly lensed. Otherwise, the particles will experience a large trajectory bending or even loop around the GC many times if $b$ approaches $b_c$. In this case, we say that the particles are strongly lensed. In Fig. \ref{figdphi}, we plot the bending angle formula \refer{exactgeneral} as a function of $v$ and $b_M$, which shows that the ultra-relativistic particles near the critical $b_c$ indeed can loop around the GC. We also numerically but directly solved the geodesic equations to obtain $r$ as a function of $\theta$ and found that the angle change between the incoming and outgoing rays from these solutions agrees perfectly with our master formula \refer{exactgeneral}.

\section{Gravitational lensing of weakly lensed ultra-relativistic particles \label{wlur}}

\begin{figure}[htbp]
\begin{center}
\includegraphics[width=0.3\textwidth]{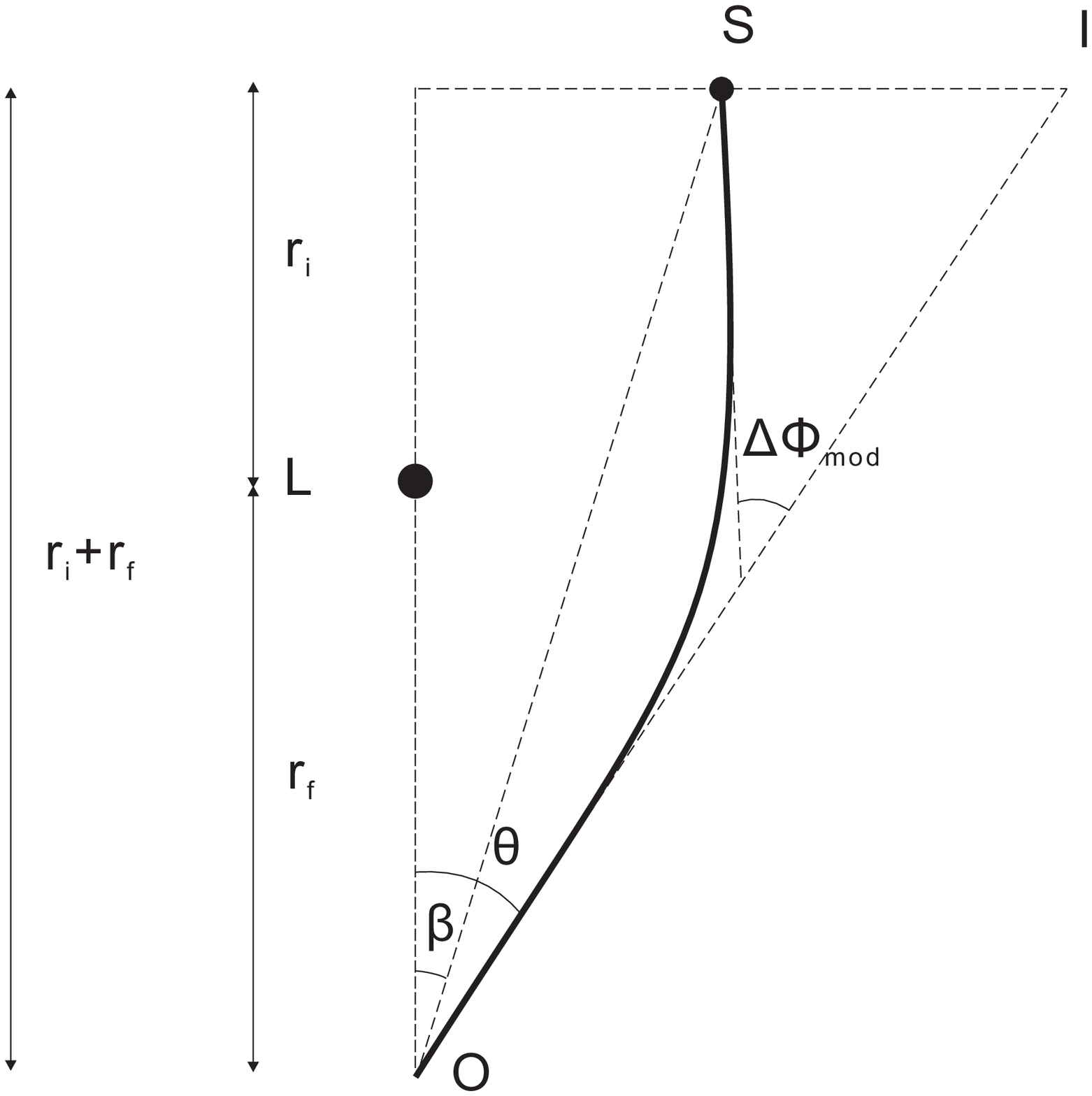}\hspace{1cm}
\includegraphics[width=0.3\textwidth]{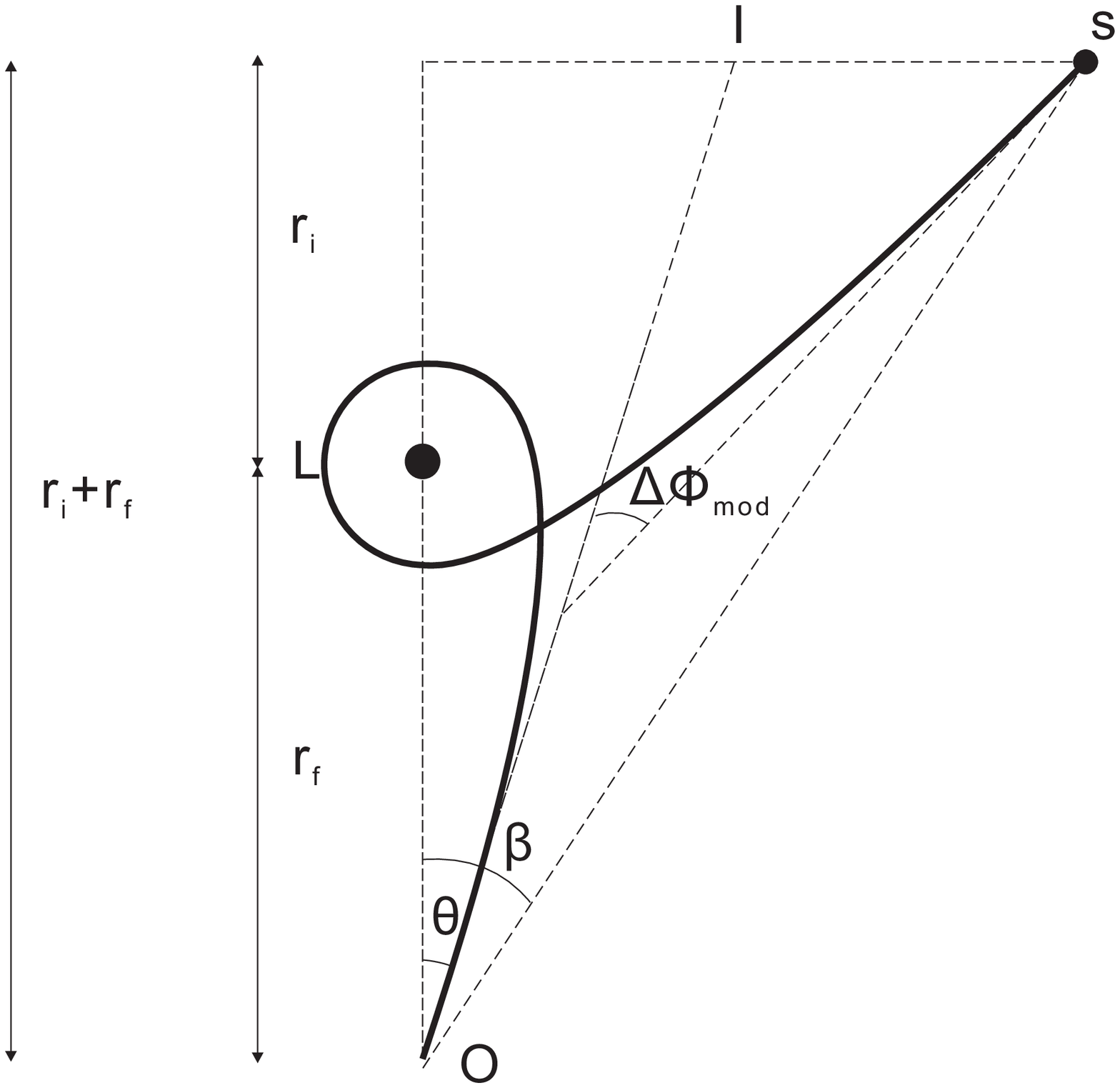}\hspace{1cm}
\includegraphics[width=0.25\textwidth]{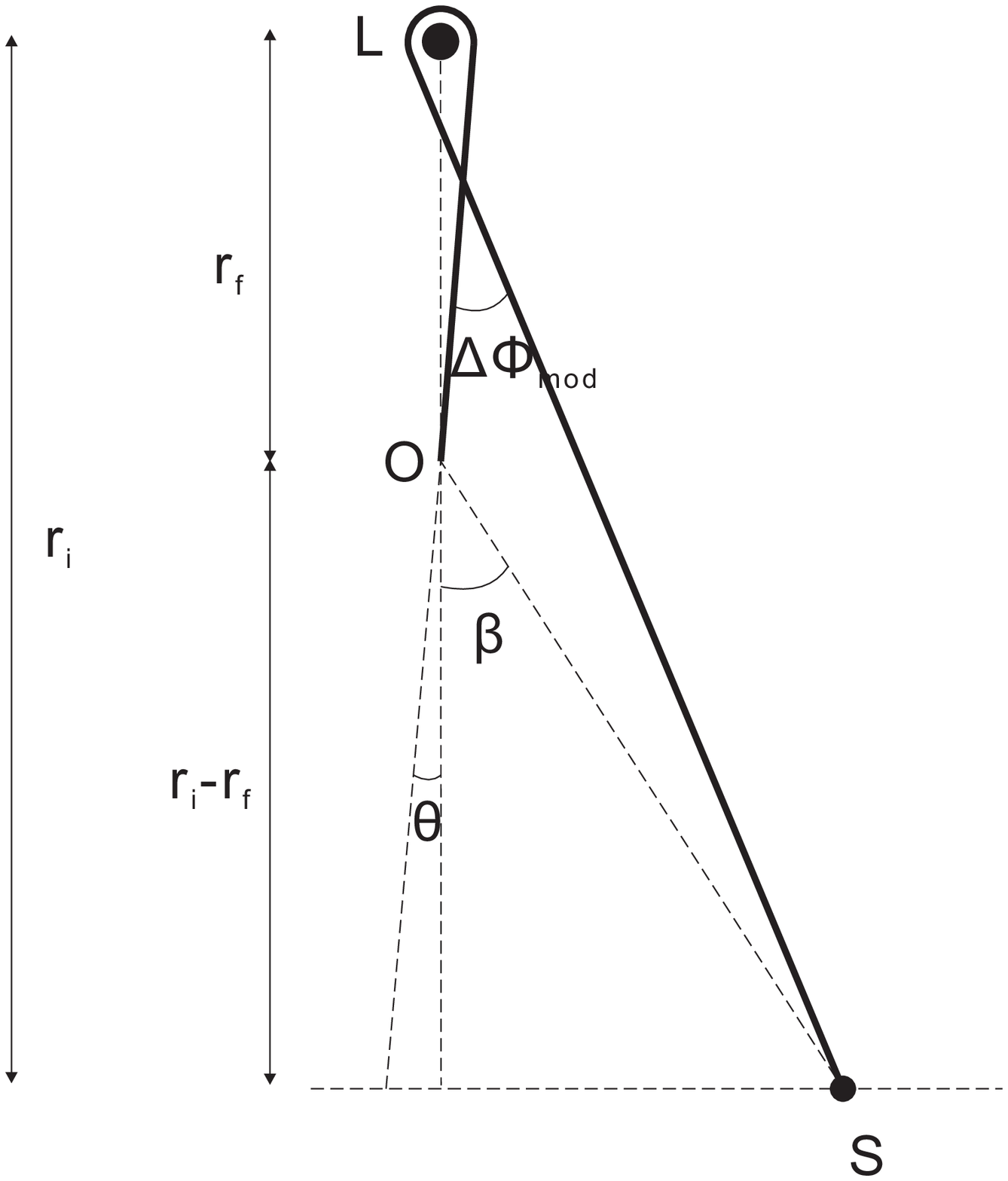}
\caption{The weak (Left) and strong (Middle) lensing with a small deflection angle for fast particles. The $b_Mv\to \infty$ limit lensing of slow particles (Right).
$\beta$ is the angle between the source $S$ and the lens-observer (LO) axis and $\theta$ is the deflection angle of the image $I$. $\Delta \phi_{\mbox{\scriptsize mod}}$ is the modulated bending angle.
}
\label{figgeom}
\end{center}
\end{figure}

For ultra-relativistic particles, as mentioned above, we can approximate the bending angle by its expansion around $v=1$ as
\be \Delta\phi(b_M,v\to 1) = \Delta\phi(b_M,1) - \Delta\phi_v(b_M,1)(1-v)+{\cal O}\left[(1-v)^2\right] . \label{v1exp}\ee
The $\Delta\phi(b_M,1)$, which is now an elliptical integral function \cite{darw2} of $b_M$ alone, yields exactly the bending of light ray given by Eq. \refer{m0ang}. The second term is typically much smaller compared to the first one for ultra-relativistic and weakly lensed (UR-WL) particles. For these particles, the classical results accumulated over the years for weakly lensed light rays, such as the deflection angle $\theta$ and the flux magnification $\mu$, can be applied approximately. We now summarize some of these results using the geometric setup in Fig. \ref{figgeom}(Left).

We can establish the general lens equation by inspecting the geometry of Fig. \ref{figgeom}(Left) as
\be (r_i+r_f)\tan\theta=(r_i+r_f)\tan\beta + r_i \Delta\phi_{\mbox{\scriptsize mod}}. \label{geolens}\ee
This relation is correct as long as all relevant angles in Fig. \ref{figgeom}(Left) are small, which is equivalent to require $r_i\gtrsim r_f\gg b$. This relation also applies to all the cases such that the modulated bending angle $\Delta\phi_{\mbox{\scriptsize mod}}$ is small, including those with $n\neq 0$ or $v\neq 1$.

For the UR-WL ray passing through a Schwarzschild gravitational field, to the lowest order,
\be
\Delta\phi_{\mbox{\scriptsize mod}}=4M/b\simeq4/b_M, \label{wlang}\ee
and together with another geometric relation
\be r_f\theta=b,\label{georel2}\ee
the lens equation \refer{geolens} can be simplified to
\be \beta=\theta-\frac{\theta_E^2}{\theta}, \mbox{ where }  \theta_E=\sqrt{\frac{4M{r_i}}{r_f(r_i+r_f)}} \label{leqsim}.\ee
This equation allow two solutions, corresponding to one primary and one secondary image, given by (see example \cite{nunthesis})
\be
\theta_{\pm}=\frac{1}{2}\left(\beta\pm\sqrt{\beta^2+4\theta_E^2}\right). \label{defang}
\ee
This result restores the dependence of the deflection angle on the initial radius $r_i$ and final radius $r_f$ of the particles. If the particle source, lens and the observer are exactly aligned, i.e. $\beta=0$, the images of the source will form a ring with radius $\theta_E$, the so called Einstein ring. When $\beta\neq0$, the two images are separated by the angle
\be \theta_+-\theta_-=\sqrt{\beta^2+4\theta_E^2} .\ee

The magnifications of these two images are given respectively by
\bea
\mu_\pm&\equiv& \left|\frac{\theta_\pm}{\beta}\frac{\dd \theta_\pm}{\dd\beta}\right|\label{magdef}\\
&=&\frac{1}{4} \left(\frac{\beta}{\sqrt{\beta^2+4\theta_E^2}}+\frac{\sqrt{\beta^2+4\theta_E^2}}{\beta}\pm2\right). \label{magni}\eea
It is seen that when $\beta$ is close to zero, the magnification of both images are large. And $\mu_+>1$ for all $\beta$ while $\mu_->1$ only when $\beta/\theta_E<\sqrt{3/\sqrt{2}-2}$. When $\beta/\theta_E$ is large, we have that $\mu_+\simeq 1$ and $\mu_-\simeq 0$.

\section{Gravitational lensing of strongly lensed ultra-relativistic particles  \label{slur}}

The above summarize the results related to the bending angle of particles that are ultra-relativistic and weakly lensed. However our continuity of the bending angle around $v=1$ is also applicable for strongly lensed (SL) ultra-relativistic particles. Therefore, the results about the deflection angle and magnification etc of strongly lensed light ray should also be applicable to SL-UR particles. These quantities for light ray were studied in Ref. \cite{Bozza:2001xd}. Here we present our own analysis for the light ray case and moreover the SL-UR particle case.

For the SL-UR particles, we still consider a near-alignment geometry except now the particles can loop around the GC (see Fig. \ref{figgeom}(Middle)). In this case, the relation \refer{geolens} still holds because the angles are still small, but now the $n$ in $\Delta\phi_{\mbox{\scriptsize mod}}$ given by Eq. \refer{exactgeneral} takes non-zero integer values $n$. Even for the lowest $n=1$, such a $\Delta\phi_{\mbox{\scriptsize mod}}$ requires that $b_M$ in Eq. \refer{exactgeneral} to approach $b_{Mc}=3\sqrt{3}$. Therefore we can expand the $\Delta\phi_{\mbox{\scriptsize mod}}$ around $b_M=b_{Mc}+\delta b_M$. To the leading order, the expansion takes the form
\be \Delta\phi_{\mbox{\scriptsize mod}}=-\ln\frac{(2+\sqrt{3})^2(b_M-b_{Mc})}{648\sqrt{3}} -(2n+1)\pi . \label{dphiexpf}\ee
For the near alignment geometry, $\Delta\phi_{\mbox{\scriptsize mod}}$ is small and then $n=1$ requires $\delta b_M\simeq 0.0065\ll b_{Mc}$, which shows that the consistency of the expansion.
Using the relation \refer{georel2}, the modulated bending $\Delta\phi_{\mbox{\scriptsize mod}}$ can be related to the deflection angle $\theta$ in the diagram using a very simple expression
\hspace{-1cm}\bea
\Delta\phi_{\mbox{\scriptsize mod}}&=&-\ln(r_f\theta/M -3\sqrt{3}) +\ln[648(7\sqrt{3}-12)]-(2n+1)\pi \nonumber\\
&\equiv&-\ln(r_f\theta/M -3\sqrt{3})+B \equiv-\ln(r_f\theta/M -3\sqrt{3})+A-2n\pi,\label{cbdef}\eea
where definitions of $B$ and $A$ are apparent.
Substituting this into Eq. \refer{geolens}, one can get a SL version of the lens equation
\be \beta=\theta-\frac{r_i}{r_i+r_f}\Delta\phi_{\mbox{\scriptsize mod}}(\theta) .\label{sleq}\ee
This will be the starting point for derivations of the SL results.

All angles that are used in Eq. \refer{sleq} have been assumed positive and the winding was assumed anticlockwise. If the winding was clockwise, we would have $\theta<0$. The size of $\theta$ can be also be solved from a lens equation similar to \refer{sleq} by substitution $\beta\to -\beta$ because there is a strict axial symmetry along the lens axis. Therefore for any $n\geq1$ and fixed $\beta$, Eq. \refer{sleq} allows two deflected images, distributed on different side of the lens axis. It is not hard to show that the deflection angle of these two images will be different although the difference is actually small for $n\geq1$.

The lens equation \refer{sleq} can be solved after using Eq. \refer{cbdef} to obtain the deflection angle as
\be
\theta_\pm(n,\beta)=\pm\left[\frac{r_i}{r_i+r_f}\mbox{W}(g(n,\pm\beta)) +\frac{3\sqrt{3}M}{r_f}\right]\label{sldefres}\ee
with $\mbox{W}$ being the Lambert-W function, $+$ and $-$ corresponding to the images on the two sides of the lens axis and
\be g(n,\beta)=\frac{(r_i+r_f)M}{r_ir_f}\exp\left[B-\frac{(3\sqrt{3}M-\beta r_f)(r_i+r_f)}{r_ir_f}\right]. \ee
With Eq. \refer{sldefres}, the dependence of the deflection angle of all images on the order $n$ and $\beta$ can be studied. As expected from the fact that all images have impact parameter close to the critical $b_{MC}$, their corresponding deflection angles are all very close to each other, decreasing with the increase of $n$ and decrease of $\beta$. Among different SL images on the same side of the source, the $n=1$ and $n=2$ images are separated the most but still very narrowly. For example at $r_i$ and $r_f$ equating the distance from our solar system to the Galaxy center, i.e., $r_i=r_f=4.04\times10^{10}M_\odot$ and $\beta=0.1$ arcsec, the angular difference of $n=1$ and $n=2$ images can reach only $3.4\times 10^{-8}$ arcsec. While the $n=1$ images on two sides under the same $\beta$, $r_i$ and $r_f$, can reach $5.3\times 10^{-5}$ arcsec. Note that lens equation \refer{sleq} can also be solved using proper series expansion \cite{Bozza:2001xd} to get
\be \theta_\pm(n,\beta)=\pm\left(\theta_{n,0}+\frac{(\pm\beta-\theta_{n,0})(\theta_{n,0}r_f-3\sqrt{3}M) (r_i+r_f)}{(r_i+r_f)(\theta_{n,0}r_f-3\sqrt{3}M)+r_ir_f}\right) \label{slsol2}  \ee with
\be \theta_{n,0}=\frac{3\sqrt{3}+e^B}{r_f/M}. \label{thetan0res}\ee
This form of the solution is particularly useful for finding the total magnification.

The magnification of each image can be calculated from the definition \refer{magdef} to be
\be \mu_\pm(n,\beta)=\frac{\theta_\pm}{\beta} \frac{\mbox{W}(g(n,\pm\beta))}{1+\mbox{W}(g(n,\pm\beta))}. \label{magsingle}\ee
As one can expect from the physical intuition, the magnification decreases very rapidly with the increase of $n$ and $\beta$. While when the source, lens and observer are perfectly aligned, the magnification diverges. Because the images of different orders for $n\geq 1$ are typically not resolvable, only one image with total flux equaling the sum of partial contributions will be observed by observatories without enough angular resolution. Therefore, it is useful to obtain a total magnification for relativistic images on each side by summing over all orders \cite{Eiroa:2003jf}
\hspace{-2.5cm} \bea 
&&\mu_\pm(\beta)=\sum_{n=1}^\infty\mu_\pm(n,\beta)=\frac{e^A(r_i+r_f)M^2}{\beta r_f^2r_i}\nonumber\\
&&\times \left\{\frac{3\sqrt{3}}{e^{2\pi}-1}
-\frac{e^A\left[(\mp\beta r_f+3\sqrt{3}M)(r_i+r_f)-r_ir_f\right]}{(e^{4\pi}-1)r_ir_f}
- \frac{e^{2A}(r_i+r_f)M}{(e^{6\pi}-1)r_ir_f}\right\},
\eea
In obtaining this, instead of Eq. \refer{magsingle} the lens equation solution \refer{slsol2} is used.
% The formula here is obtained from arxiv. 0311013. using a_2=A-ln(2).

\section{Velocity correction to the bending angles \label{vcor1}}

When the particle is not massless, its velocity $v$ at infinity will deviate from the speed of light. Therefore there will be corrections to the bending angle \refer{exactgeneral} due to this velocity difference and consequently modifications to the lens equation \refer{geolens}. For non-relativistic $v$, this modification can even be large. Unfortunately, for general $v$ we need to directly substitute Eq. \refer{exactgeneral} into Eq. \refer{geolens}, which without further approximations will only be solvable numerically due to the complicated elliptical function form. In the two limiting cases that $v$ approaches 1 or 0 however, we can tackle the lensing equation perturbatively. In this section, we will do analytical studies of the lens equations for these two limits.

\subsection{The fast particle case\label{vcloseto1}}

The second term in Eq. \refer{v1exp} is the correction to the bending angle due to the small velocity deviation from 1. In this section, we study quantitatively the effect of this deviation to the bending angle and the deflection angle of the images.

Using Eq. \refer{exactgeneral}, the coefficient function $\Delta\phi_v(b_M,1)$  can be computed as
\be \Delta\phi_v(b_M,1)=c_EE(a,b)+c_FF(a,b)+c_0\label{dpvdef}\ee
where $E$ and $F$ are the incomplete elliptical integral of the second and first kind respectively, and
\hspace{-1cm}\bea
a&=&\frac{\sqrt{w_1}}{\sqrt{w_1-w_2}},~b=\frac{\sqrt{w_1-w_2}i}{\sqrt{w_3-w_1}},\\
c_0&=&-\frac{4b_M}{b_M^2-27}, ~c_E=\frac{72\sqrt{3}}{b_M(b_M^2-27)^{3/4}\sqrt{1+2\cos(2x/3)}},\\
c_F&=&\frac{c_E b_M}{18}\left\{\sin\left(\frac{2x}{3}+y\right) -\sin\left(\frac{x}{3}+y\right)  +\frac{\left(\sqrt{b_M^2-27}-9\right)}{f}\right\},\\
x&=&\tan^{-1}\left(\frac{6\sqrt{3b_M^2-81}}{b_M^2-54}\right)-\mbox{Heaviside}(b_M^2-54)\pi, ~y=\tan^{-1}\left(\frac{\sqrt{3b_M^2-81}}{9}\right).
\eea
Here $w_1$ and $w_2$ are given by Eqs. \refer{w1fv} at $v=1$ and $w_3$ is the larger positive root of Eq. \refer{cubeq} at $v=1$ which is given by
\be w_3=\frac{1}{2}-w_1-w_2.\ee
Since no approximation regarding $b_M$ was used in the expansion \refer{v1exp}, the coefficient function $\Delta\phi_v$ we found in Eq. \refer{dpvdef}  will be applicable to all values of $b_M$, i.e., both the WL and SL cases.

For the WL case, from Eq. \refer{wlang} we know that the light ray bending decreases as $4/b_M$ when $b_M$ increases. For the SL case, from \refer{dphiexpf} we observe that the bending of light will diverge logarithmically as $b_M$ approaches $b_{Mc}$.   It would be interesting here to see what kind of $b_M$ dependence this velocity correction term will have, since it might become competitive to the leading term (the bending for $v=1$) if it grows faster or decreases slower as $b_M$ approaches the $b_{Mc}$ limit or infinity. We have carried out this analysis and found that: (1) for the WL case, i.e., when $b_M\to\infty$, the correction to the light bending angle in Eq. \refer{v1exp} takes the form
\be -\Delta \phi_v(b_M,1)(1-v)=-\left[-\frac{4}{b_M}+{\cal O}\left(\frac{1}{b_M^2}\right)\right](1-v); \label{bafastsf}\ee
and (2) for the SL case, i.e., when $b_M\to b_{Mc}$, the correction takes the form
\be \hspace{-1.5cm} -\Delta \phi_v(b_M,1)(1-v) = -\left\{-\frac{2\sqrt{3}}{\delta} +\left[14-8\sqrt{3}+\ln\left(\frac{(4\sqrt{3}+7) \delta}{648\sqrt{3}}\right)\right]/27  +{\cal O}\left(\delta^{1/2}\right)\right\}(1-v) \label{bafastlf}\ee
where $ \delta=b_M-3\sqrt{3}$.

We can compare this correction to the corresponding light ray bending angles by taking their ratios. For the WL and SL case, the ratios can be computed using Eqs. \refer{wlang} and \refer{dphiexpf}. To the leading order, they are respectively
\bea &&\frac{-\Delta\phi_v(b_M,1)(1-v)}{\Delta\phi(b_M,1)}\propto (1-v)\frac{4/b_M}{4/b_M} \stackrel[b_M\to\infty]{\longrightarrow}{} (1-v),\nonumber\\
&& \frac{-\Delta\phi_v(b_M,1)(1-v)}{\Delta\phi(b_M,1)}\propto (1-v)\frac{2\sqrt{3}/(b_M-3\sqrt{3})}{\ln(b_M-3\sqrt{3})}
\stackrel[b_M\to3\sqrt{3}]{\longrightarrow}{} \infty. \nonumber\eea
It is clear that while for the WL case the correction is usually small, the correction for the SL case might compete with the bending angle of light ray.

\begin{figure}[htbp]
	\begin{center}
%		%\includegraphics[width=0.3\textwidth]{dphidvnew.eps} %depending on the latex compiler, you can omit the file extension
		\includegraphics[width=0.4\textwidth]{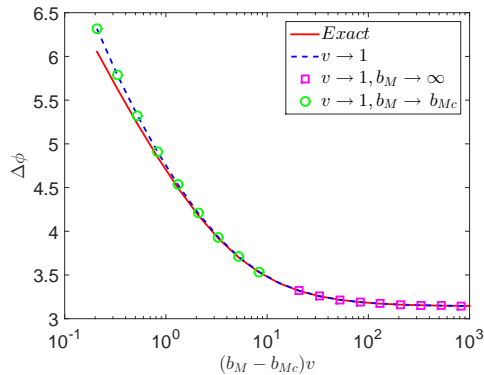} %depending on the latex compiler, you can omit the file extension
		\caption{The bending angle as a function of $b_M$ at a representative large $v=0.99$. The red solid and dash blue curves are obtained from the exact formula \refer{exactgeneral} and its large $v$ expansion \refer{v1exp} with \refer{dpvdef} respectively. The circle and square symbols are calculated from the small $b_M$ expansion \refer{bafastsf} and large $b_M$ expansion \refer{bafastlf} respectively. }
		\label{phivfig}
	\end{center}
\end{figure}

We plot in Fig. \ref{phivfig} the bending angle as a function of $b_M$ for a large velocity $v=0.99$. The exact formula \refer{exactgeneral}, the large $v$ expansion \refer{v1exp} and its two limits of $b_M$ \refer{bafastsf} and \refer{bafastlf} are all plotted. It is seen that the all agrees very well.

\subsection{The slow particle case}

When the particle ray is slow, its geodesic trajectory will not assemble that of the light ray. And since in this case, the difference between the particle velocity and light speed can not be considered as a perturbation, we will not be able to expand the bending angle around $v=1$ as done in Eq. \refer{v1exp}. On the other hand, we can not directly expand the \refer{exactgeneral} around a small velocity $v$ while holding $b_M$ constant either. The reason is precisely our former observation Eq. \refer{fcdef} that for slow particle, the $f=b_{M}v$ should be larger than a finite critical value $f_c$. Therefore what we can do is only to hold $b_Mv$ to a value larger than $f_c$ while expanding over $v$. We carried out this expansion and found that it takes the form
\be
\Delta\phi(b_M, v\to 0)=\Delta\phi(b_M,v=0)+\Delta\phi_{v^2}(b_M,0)v^2+{\mathcal O}(v^4) \label{baslow}.\ee
Here the first term
\be \Delta\phi(b_Mv\neq 0,v=0)
=\sqrt{f}(\sqrt{f+4}-\sqrt{f-4})K(l),\ee
with $f=b_Mv$, $K$ being the complete elliptic integral of the first kind and $ l=(f-\sqrt{f^2-16})/4 .$
The second term is given by
\bea
\Delta\phi_{v^2}(b_M,0)v^2 &=&\left\{
\frac { ( f^2-14 ) f^{3/2} ( \sqrt {f+4}+\sqrt {f-4} ) }{2(f^2-16)}E(l) \right.\nonumber\\
&&\left.-\frac{\sqrt{f}}{2}\left(\frac{f^2-3f-2}{\sqrt{f-4}}+\frac{f^2+3f-2}{\sqrt{f+4}}\right)K(l) \right\}v^2. \label{bfsvexp2ndterm}
\eea

\begin{figure}[htbp]
	\begin{center}
%		%\includegraphics[width=0.3\textwidth]{dphidv.eps} %depending on the latex compiler, you can omit the file extension
		\includegraphics[width=0.4\textwidth]{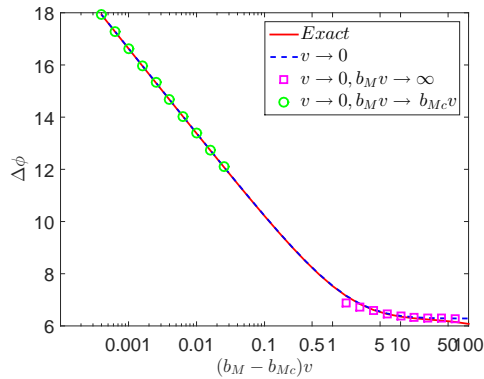} %depending on the latex compiler, you can omit the file extension
		\caption{\label{phiv0fig} The bending angle as a function of $b_M$ for a small velocity $v=10^{-3}$.  The red solid and dash blue curves are obtained from the exact formula \refer{exactgeneral} and its small $v$ expansion \refer{baslow} respectively. The circle and square symbols are calculated from the small $b_Mv$ expansion \refer{svsbexp} and large $b_Mv$ expansion \refer{svlbexp} respectively. }
	\end{center}
\end{figure}

As in the case of fast particle, we can further expand the angle \refer{baslow} for the two limiting case of $b_M$: (1) the weak lensing case $b_M v \to \infty$ and (2) the strong lensing case $b_M\to b_{Mc}=4/v$. We found that for the WL case, the $\Delta\phi$ takes form
\be \hspace{-1cm}
\Delta\phi(b_Mv\to \infty,v\to 0)= 2\pi\left(1+\frac{3}{b_M^2v^2}\right) +\frac{3\pi}{2b_M^2v^2}\left(1+\frac{35}{b_M^2v^2}\right)v^2+\mbox{higher orders},\label{svlbexp}\ee
and for the SL case,
\bea\hspace{-2cm}
\Delta\phi(b_Mv\to 4,v\to 0)&=& \frac{\sqrt{2}\left[(32\delta^{-1} +3)\ln(128\delta^{-1}) +2\right]}{32\delta^{-1}} +\frac{\sqrt{2}(32\delta^{-1} -27\ln(128\delta^{-1}) +137)}{16} v^2 \nonumber\\
&&+\mbox{higher orders},\label{svsbexp}\eea
where $\delta^{-1}\equiv (b_Mv-4)^{-1}$ is large when $b_Mv\to 4$.
For the WL case, it is seen that the subleading term decreases faster than the leading term  and therefore the leading term will dominate. For the SL case, as $b_M$ approaches the $b_{Mc}$, both the leading term and the subleading term will increase. However, the subleading term dose this more rapidly and will become competitive to the leading term eventually.  Putting contributions of different orders of $v$ into a ratio form, we have
\bea
&&\frac{\Delta\phi_{v^2}(b_M,0)v^2}{\Delta\phi(b_M,0)} \stackrel[b_Mv\to\infty]{\propto}{} \frac{3\pi /(2b_M^2)}{2\pi}\to0,\label{wlslow}\\
&&\frac{\Delta\phi_{v^2}(b_M,0)v^2}{\Delta\phi(b_M,0)}\stackrel[b_Mv\to 4]{\propto}{} \frac{2\sqrt{2}\delta^{-1}v^2}{\sqrt{2}\ln(\delta^{-1})}\to\infty. \label{slslow}
\eea
Therefore when using the expansions \refer{svsbexp} one has to make sure that the $b_M$ and $v$ are such that the second and higher order in $v$ terms will not dominate the result of leading order in $v$.

It is worth noticing that for the WL case, as $b_M$ approaches infinity, the bending angle approaches $2\pi$ from above. This suggests that for any particle with low speed $v$ but large impact parameter $b$ (such that $b_Mv$ is also large), the particle will always travel backward along almost the same angular direction. If these particles can be gravitationally lensed, then it will form a retro-lensing situation \cite{Holz:2002uf, Eiroa:2003jf}. Note that in Newtonian mechanics, the bending angle of a particle with velocity $v$ at infinity takes the form
\be \Delta\phi=2\pi-2\cos^{-1}\left[1+v^8b_M^2/4\right] \ee
whose low speed limit yields
\be \Delta\phi=2\pi-b_Mv^4+{\mathcal O}(v^{12}) .\ee
This bending also approaches $2\pi$ but from below. This is apparently different from the GR bending angle \refer{svlbexp} but understandable from the known facts that for a particle with positive energy at infinity, its Newtonian orbit is a hyperbola and therefore never crosses itself and the bending angle never excesses $2\pi$.

The bending angle as a function of $b_M$ for a particle with small velocity $v=10^{-3}$ has been plotted in Fig. \ref{phiv0fig}. The result obtained using the exact formula \refer{exactgeneral}, the small velocity expansion \refer{baslow} and its two expansions \refer{svsbexp} and \refer{svlbexp} are plotted. It is seen that our expansion in $v$ agrees very well with the exact result, and both of them agree very well with the corresponding $b_Mv$ limits.

\section{Velocity correction to the deflection angles}

With the above correction due to velocity to the bending angle $\Delta\phi$ in hand, in this section we will examine the corresponding lens equations to study the effect of velocity to the deflection angles. Again, we separately discuss the cases of fast particle and slow particles.

\subsection{Deflection angle of fast particles}

As in the case of light rays, we study the deflection angle of fast particles according to whether they are weakly of strongly lensed. For weakly lensed particles, we started from the lens equation \refer{geolens} and for fast particle, we used the bending angle \refer{wlang}, which to the lowest order is just the bending of light ray. To study the correction due to velocity deviation from the speed of light to the deflection angle, we can directly substitute Eqs. \refer{bafastsf} and \refer{wlang} into Eq. \refer{geolens}. The resultant equation becomes
\be \beta=\theta-\frac{\theta_E^{\prime 2}}{\theta}, \mbox{ where }  \theta_E^\prime=\sqrt{\frac{4(2-v)Mr_i}{r_f(r_i+r_f)}} \label{leqsimfw},\ee
whose solution gives the deflection angle
\be
\theta_{\pm}^\prime=\frac{1}{2}\left(\beta\pm\sqrt{\beta^2+4\theta_E^{\prime 2}}\right). \label{defangfw}
\ee
Comparing to the original deflection angle \refer{defang}, we see that the correction is about
\be \frac{\theta_\pm^\prime-\theta_\pm}{\theta_\pm}\approx\frac{8Mr_i(1-v)}{C \pm\beta\sqrt{Cr_f(r_i+r_f)}}, \ee
where $C=\beta^2 r_f(r_i+r_f)+16Mr_i$.

For strongly lensed particles, our correction to the deflection angle should start from
the Eq. \refer{sleq}. The $\Delta\phi_{\mbox{\scriptsize mod}}$ here is the small angle obtained from the combination of Eqs. \refer{cbdef} and \refer{bafastlf}. If we directly solve the resultant equation for the deflection angle $\theta$ however, we will run into a more complicated formula than Eq. \refer{sldefres}. Therefore here we follow the strategy of Ref. \cite{Bozza:2001xd} by find a series expansion for the $\Delta \phi_{\mbox{\scriptsize mod}}(\theta)$ term in Eq. \refer{sleq}. We first solve the values $\theta^\prime_{n,0}$ for different indices $n$, around which we will do the expansion.  Using Eqs. \refer{cbdef} and \refer{bafastlf} we get
\be 0=\Delta\phi_{\mbox{\scriptsize mod}}(\theta_{n,0}^\prime)
=-\ln(r_f\theta_{n,0}^\prime/M-3\sqrt{3})+B +\frac{2\sqrt{3}}{r_f\theta_{n,0}^\prime/M-3\sqrt{3}}(1-v) \ee
from which we obtain $\theta_{n,0}^\prime$, the counterpart of $\theta_{n,0}$ in Eq. \refer{thetan0res} but with velocity correction, as
\be
\theta_{n,0}^\prime= \frac{\sqrt{3}M}{r_f}\left(3+\frac{2(1-v)}{\mbox{W}(g^\prime(n,v))} \right) \label{theta0n}\ee
where W is the Lambert-W function and
\be g^\prime(n,v)=\frac{(7+4\sqrt{3})(1-v)}{324} e^{(2n+1)\pi}.\ee
When $v$ is close to 1, this becomes
\be \theta_{n,0}^\prime =\theta_{n,0} +\frac{2\sqrt{3}(1-v)}{r_f}, \ee
where $\theta_{n,0}$ is the Eq. \refer{thetan0res} and the second term is its correction.
Then the $\Delta\phi_{\mbox{\scriptsize mod}}(\theta)$ can be expanded around $\theta_{n,0}^\prime$ and after substituting into Eq. \refer{sleq} we can solve the deflection angle, which is roughly the same form as Eq. \refer{slsol2} but with $\theta_{n,0}$ replaced by $\theta_{n,0}^\prime$
\be \theta_\pm^\prime(n,\beta)
\approx \pm\left(\theta_{n,0}^\prime+\frac{(\pm\beta-\theta_{n,0}^\prime)(\theta_{n,0}^\prime r_f-3\sqrt{3}M) (r_i+r_f)}{r_ir_f+2\sqrt{3}r_ir_f(1-v)/(\theta_{n,0}^\prime -3\sqrt{3}M)}\right) . \ee
Again, since $v$ is close to 1 and $r_i,~r_f$ are large, we can get through series expansion a simplified result
\be \theta_\pm^\prime(n,\beta)\approx\theta_\pm(n,\beta)
+\frac{2\sqrt{3}M}{r_f}\frac{r_i}{(r_i+r_f)\theta_{n,0}+r_i}(1-v). \ee
The second term here apparently is the correction of velocity to the deflection angles of relativistic images in the SL case. It is seen that besides the expected proportional factor $(1-v)$, the correction receives an extra suppression due to the first factor $2\sqrt{3}M/r_f$ which is usually very small since $r_f$ is large.

\begin{table*}[htp]
\small
\centering
\begin{tabular}{ccc|c|c|ccc}   % v=1
\hline
$\theta_-(2,-\beta)$&$\theta_-(1,-\beta)$&$\theta_-(0,-\beta)$&$\beta$&$v$&$\theta_+(0,+\beta)$&$\theta_+(1,+\beta)$&$\theta_+(2,+\beta)$\\
\hline
\hline
$-2.6537\times10^{-5}$&$-2.6571\times10^{-5}$&$-1.451$&$10^{-3}$&$1$&$1.452$&$2.6571\times10^{-5}$&$2.6537\times10^{-5}$\\
$-2.6714\times10^{-5}$&$-2.6748\times10^{-5}$&$-1.458$&$10^{-3}$&$0.99$&$1.459$&$2.6748\times10^{-5}$&$2.6714\times10^{-5}$\\
$-2.6537\times10^{-5}$&$-2.6571\times10^{-5}$&$-1.447$&$10^{-2}$&$1$&$1.457$&$2.6571\times10^{-5}$&$2.6537\times10^{-5}$\\
$-2.6714\times10^{-5}$&$-2.6748\times10^{-5}$&$-1.454$&$10^{-2}$&$0.99$&$1.464$&$2.6748\times10^{-5}$&$2.6714\times10^{-5}$\\
$-2.6537\times10^{-5}$&$-2.6571\times10^{-5}$&$-1.402$&$10^{-1}$&$1$&$1.502$&$2.6571\times10^{-5}$&$2.6537\times10^{-5}$\\
$-2.6714\times10^{-5}$&$-2.6748\times10^{-5}$&$-1.409$&$10^{-1}$&$0.99$&$1.509$&$2.6748\times10^{-5}$&$2.6714\times10^{-5}$\\
$-2.6537\times10^{-5}$&$-2.6571\times10^{-5}$&$-1.035$&$1$&$1$&$2.035$&$2.6571\times10^{-5}$&$2.6537\times10^{-5}$\\
$-2.6714\times10^{-5}$&$-2.6748\times10^{-5}$&$-1.042$&$1$&$0.99$&$2.042$&$2.6748\times10^{-5}$&$2.6714\times10^{-5}$\\
$-2.6537\times10^{-5}$&$-2.6571\times10^{-5}$&$-0.2064$&$10$&$1$&$10.206$&$2.6571\times10^{-5}$&$2.6537\times10^{-5}$\\
$-2.6714\times10^{-5}$&$-2.6748\times10^{-5}$&$-0.2084$&$10$&$0.99$&$10.208$&$2.6748\times10^{-5}$&$2.6714\times10^{-5}$\\
\hline
\end{tabular}\\
\caption{The deflection angles $\theta_\pm(n,\pm \beta)$ with indices $n=0, ~1$ and $2$ for some representative $\beta$. We assume that the source is located on the opposite side of the Galaxy with the same distance from the central black hole as the solar system, so that $r_i=r_f\simeq 8.33$ kpc. We used $M=4.31\times 10^6M_{\odot}$ \cite{Gillessen:2008qv}. Unit of angles is arcsec. \label{twlda}}
\end{table*}

In Table \ref{twlda}, the numerical values of the deflection angles for a source located on the opposite side of the Galaxy with respect to the solar system with the same distance from the central black hole are present. We give the results for both light rays and particles with velocity $v=0.99$. It is seen that the primary and secondary images, denoted by $\theta_+(0,\beta)$ and $\theta_-(0,\beta)$, are most affected by $\beta$ among all images. The relativistic images with different $n$ are separated very narrowly and therefore they are beyond the resolution of current and near future observatories for light rays or other very fast particles.
\vspace{5mm}

\begin{figure}[htbp]
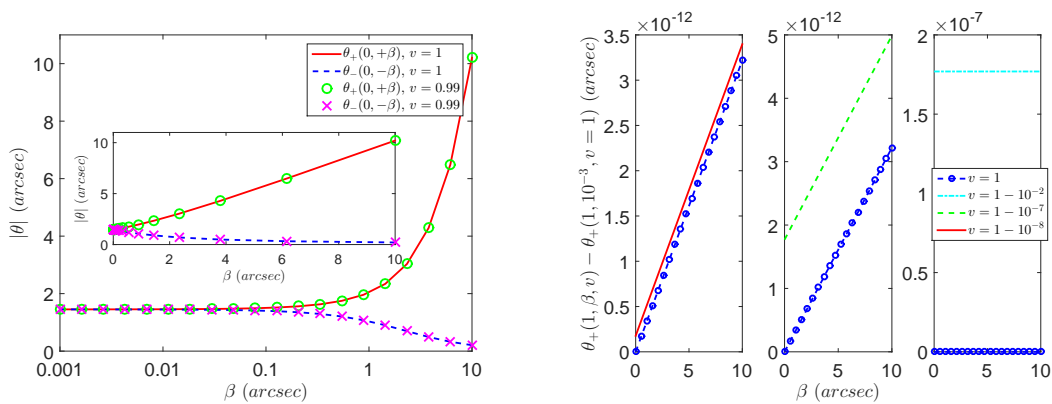

	\begin{center}
%		%\includegraphics[width=0.3\textwidth]{dphidv.eps} %depending on the latex compiler, you can omit the file extension
		\includegraphics[width=0.4\textwidth]{deflangle_betaWL_20151213T183606.eps} \hspace{1cm}
		\includegraphics[width=0.4\textwidth]{deflangle_betaSL_20151213T183849.eps}
		\caption{\label{defangfig} The deflection angles for $\beta$ from $10^{-3}$ to $10$ arcsec and $v=1$ and $v=0.99$.  Top: (a) The deflection angle of the primary and secondary images. The inserted figure is a linear plot. Bottom: (b) The deflection angles of the relativistic image with index $n=1$ for different velocities. Other parameters are the same as in Table \ref{twlda}.}
	\end{center}
\end{figure}

The corrections to the deflection angles for different $\beta$ and $v$ in the WL and SL cases are plotted in Fig. \ref{defangfig}. Fig. \ref{defangfig} (a) shows the dependence of the deflection angles of the primary and secondary images on the source position $\beta$ and the particle velocity. It is seen that for a typical range of $\beta$ from $10^{-3}$ to $10^1$ arcsec and $v$ from $0.99$ to 1, the variation in the deflection angles are mainly caused by the change of $\beta$ rather than by change of $v$. It is also evident that the primary and secondary images are almost symmetric with respect to the lens axis when $\beta$ is smaller than roughly $1$ arcsec, while this symmetry is lost for larger $\beta$. The inserted figure shows that image deflection angle for all considered $v$ is roughly a linear function of the source position $\beta$. Fig. \ref{defangfig} (b) shows the deflection angle of relativistic images with $n=1$ for different $v$ and $\beta$. Unlike the primary and secondary images, the relativistic image deflection angle $\theta^\prime_+(1, \beta)$ can depend on $v$ more sensitively once $v$ deviates from 1 by a value larger than $10^{-7}$ (the middle panel of Fig. \ref{defangfig} (b)). If the deviation is below this value, the change of the deflection angle caused by variation of $\beta$ will be much larger than that caused by $v$ (left panel). While above this value, the opposite happens (right panel). Moreover, comparing Fig. \ref{defangfig} (a) and (b) one can also observe that the correction to the angles due to $v\neq 1$ decreases as $n$ varies from $0$ to $1$. All these corrections are typically small compared to the deflection angles themselves.

\subsection{Deflection angle of slow particle}

Like the case of fast particles, we can study the deflection of slow particles for the weak and strong lensing cases. However, unlike the case of fast particle, we have to be careful when using the slow particle bending angles together with the lens equation \refer{geolens}. The reason is that, different from the fast particle case, the slow particle bending angles \refer{svlbexp} and \refer{svsbexp} are only valid for $b_Mv\to \infty$ and $b_M v\to 4$ respectively, but not simply for $b_M\to\infty$ or $b_M\to b_{Mc}$. Both these limits requires the impact parameter to approach a large value when $v\to 0$. When $b_M$ is large comparing to the source-lens distance $r_i$ or the lens-observer distance $r_f$ however, both the lensing geometries in Fig. \refer{figgeom} will be broken and therefore the lens equation will not be applicable. Therefore we restraint our following discussion in this section to the case that $b_M$ is large but still much smaller than $r_i$ and $r_f$.

As pointed out in last section, for slow particle with large $b_Mv$, the bending angle is close to $2\pi$ and therefore only retro-lensing can occur. A schematic diagram for retro-lensing is plotted in Fig. \ref{figgeom}(Right). One can establish a geometric relation
\be r_i\Delta\phi_{\mbox{\scriptsize mod}}=(\theta-\beta)(r_i-r_f) .\label{retrogeorel}\ee
Here $\beta$ is defined as the polar angle from the observer-lens line to the observer-source line modulating $\pi$ so that $|\beta|\ll 1$. Consequently $\beta$ in Fig. \ref{figgeom}(Right) will be negative and this way the geometric relation \refer{retrogeorel} will hold even when $r_f$ is larger than $r_i$.

Substituting Eq. \refer{svlbexp} in and using Eq. \refer{georel2} we obtain a cubic equation of the deflection angle $\theta$
\be \theta^3-\beta\theta^2+\frac{6\pi r_iM^2}{r_f^2(r_f-r_i)v^2}=0. \ee
Only one real solution exists for this equation:
\be \theta =\frac{z}{6}+\frac{2\beta^2}{3z}+\frac{\beta}{3}, \label{svlfdefang}\ee
where
\be z= \left(-108d+8\beta^3+12\sqrt{-12\beta^3d+81d^2}\right)^{1/3},~
d=\frac{6\pi r_iM^2}{r_f^2(r_f-r_i)v^2}. \ee

\begin{figure}[htbp]
\begin{center}
\includegraphics[width=0.4\textwidth]{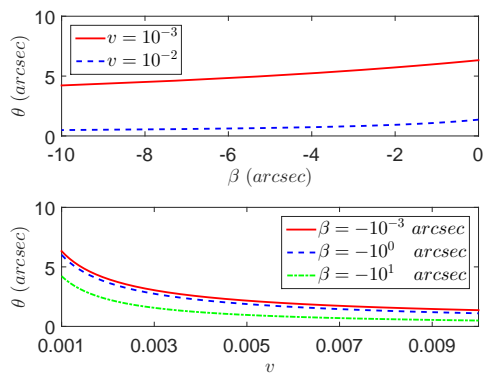}
\caption{\label{retrofig} The deflection angles of gravitationally lensed slow particles with large $bv$ for $\beta$ from $10^{-3}$ to $10$ arcsec (top panel) and $v$ from $10^{-3}$ to $10^{-1}$ (bottom panel). We assume that the source is located on the edge of the Galaxy stellar disk and therefore $r_i=13.9$ kpc \cite{Minniti:2011ag}. Other parameters are set as the same as in Table \ref{twlda}.}
\end{center}
\end{figure}

To study this deflection angle, we take an example of particles that originate from the edge of our Galaxy stellar disk and then retro-lensed to us by the Galaxy center. In Fig. \ref{retrofig}, we plotted the deflection angle as a function of $\beta$ and $v$. A close look at the top panel shows that unlike fast particle case, the dependence of deflection angle on $\beta$ is not very linear anymore. To understand this behavior, we inspected  Eq. \refer{svlfdefang} and found that in the ranges of parameters $(r_i, ~r_f, \beta, ~v)$ considered, all three terms are of similar size. And they are such that when combined, they show a deviation from a linear relation. As for the dependence on $v$, first it is noticed that as $v$ decrease, the deflection angle becomes larger. In fact, this feature is quite natural: one would expect that for a slower particle, it should have shot with a larger $b$ in order to reach the same observer and therefore will appear with a larger $\theta$ to the observer. Finally it is observed that the in the ranges of parameters considered, the changes of the deflection angle caused by change of $v$ and change of $\beta$ are of similar size: both are a few arcsecs.

For slow particles that are strongly lensed, from Eq. \refer{svsbexp} we see that particle rays will loop around. Therefore for this case we will solve a usual lensing as shown in the schematic diagram Fig. \ref{figgeom}(Middle). In this case, Eq. \refer{geolens} still holds and our modulated bending angle can be deduced from Eq. \refer{svsbexp}. We see that in this equation the leading contribution to $\Delta\phi$ is given by the $\ln(128\delta^{-1})$ term, therefore
\be \Delta\phi_{\mbox{\scriptsize mod}}=\sqrt{2} \ln[128(b_Mv-4)^{-1}]-(2n+1)\pi. \ee
Substituting this into \refer{geolens} and using Eq. \refer{georel2}, we can obtain a SL equation for the slow particle case, which formally is the same as Eq. \refer{sleq}.
Like in the SL of fast particle case, we can solve this resultant equation in two ways. First we can directly solve the deflection angle involving a Lambert-W function
\be
\theta_{\pm}(n,\beta)=\pm\left[\frac{\sqrt{2}r_i}{r_i+r_f} \mbox{W}(k(n,\pm \beta,v))+\frac{4M}{r_fv}\right]\ee
where
\be k(n,\beta,v)=\frac{64\sqrt{2}(r_i+r_f)M}{r_ir_fv} \exp \left[\frac{1}{\sqrt{2}}\left(\frac{(r_i+r_f)(\beta r_f v-4)}{r_ir_fv}-(2n+1)\pi\right)\right].\label{dasp1}\ee
Note that in this case the deflection angle of relativistic images becomes a function of not only $\beta$ but the particle velocity $v$.
We can also solve the lens equation using a series expansion of the deflection angle around angles $\theta=\theta^{\prime\prime}_{n,0}$ that let $\Delta\phi_{\mbox{\scriptsize mod}}=0$, i.e.,
\be 0=\Delta\phi_{\mbox{\scriptsize mod}}=\sqrt{2} \ln[128(\theta^{\prime\prime}_{n,0} r_f v-4)^{-1}]-(2n+1)\pi. \ee
From this we obtain
\be \theta^{\prime\prime}_{n,0}(v)=\frac{128\exp\left(-\frac{(2n+1)\pi}{\sqrt{2}}\right)+4}{r_fv/M}. \label{tppn0}\ee
Then substituting this into the lens equation and make series expansion of $\theta$ around $\theta^{\prime\prime}_{n,0}$ we obtain
\be \theta_{\pm}(n,\beta,v)=\pm\left[\theta^{\prime\prime}_{n,0} +\frac{(\pm\beta-\theta^{\prime\prime}_{n,0})(\theta^{\prime\prime}_{n,0}r_fv-4M)(r_i+r_f)} {(r_i+r_f)(\theta^{\prime\prime}_{n,0}r_fv-4M)+\sqrt{2}r_ir_fv}\right].\label{dasp2}\ee
Here the second term on the right hand side is actually small comparing to the first since it is as a perturbation to $\theta^{\prime\prime}_{n,0}$.

\begin{figure}[htbp]
	\begin{center}
%		%\includegraphics[width=0.3\textwidth]{dphidv.eps} %depending on the latex compiler, you can omit the file extension
		\includegraphics[width=0.4\textwidth]{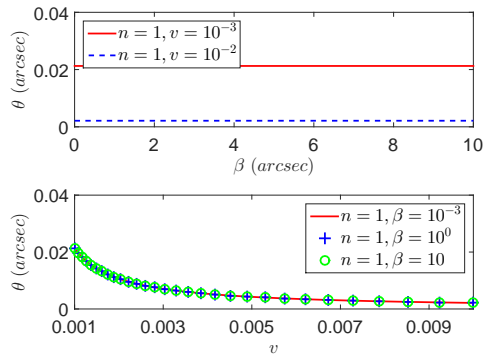} %depending on the latex compiler, you can omit the file extension
		\caption{\label{slownormalfig} The deflection angles for strongly lensed slow particles for $\beta$ from $10^{-3}$ to $10$ (top panel) arcsec and $v$ from $10^{-3}$ to $10^{-1}$ (bottom panel). We show the values for relativistic image with index $n=1$. Other parameters are set as the same as in Table \ref{twlda}.}
	\end{center}
\end{figure}

Fig. \ref{slownormalfig} shows the deflection angles \refer{dasp2} as a function of $\beta$ and $v$. In the denominators of both Eqs. \refer{tppn0} and \refer{dasp2}, we see that there exist a $v$ and therefore it is expected that the deflection angle will roughly be proportional to $1/v$. This is indeed the case as can be seen from Fig. \ref{slownormalfig} bottom panel. This feature can be similarly understood as in the case of lensing of slow particle with large $b_Mv$: a smaller $v$ requires larger $b$ and therefore results in a larger $\theta^{\prime\prime}_{n,0}$.  As for the dependence on $\beta$, it is seen that this slow particle case behaves just like the fast particle case (see Fig. \ref{defangfig}) - both are roughly linear in the range of $\beta$ we considered. As a benchmark test, we numerically calculated $\theta^{\prime\prime}_{n,0}r_fv$ using the values in this figure because we know that by approximation $f=b_Mv=\theta^{\prime\prime}_{n,0}r_fv$ should approach the limit $f_c(v\to 0)=4$ in this strong lensing case. We found that these values of $\theta^{\prime \prime}_{n,0}$ do satisfy this limit and this shows the correctness of the computations. Finally one also see that the relativistic image deflection angles depend much more strongly on $v$ than on $\beta$. Indeed, this is also expected from the physical point of view - after all, relativistic images are from rays that are close to the particle sphere whose radius approaches to $4/v$ when $v$ is small.

\section{Discussion\label{discus}}

The application of the bending angles and GL of non-zero mass objects relies on the simple fact that the bending angle and deflection are determined by the GC, the object impact parameter and velocity. Consequently the study/observation of the GL of such objects will reveal information about them. We would like to point out two circumstances that our result might be relevant.

The first is in the motion of hypervelocity stars \cite{Yu:2003hj, Brown:2005ta,hills}. It is known that some mechanisms, such as BH tidal force on a binary system, or the slingshot by a thermonuclear supernova in a close binary \cite{Guillochon:2014qpa}, can eject a start with velocity greater than 300 km$\cdot$s$^{-1}$. Previously such stars has been observed and their velocities and positions are fitted to obtain information about the GC, the surrounding or other parameters in the model. From the point of view of this work, these stars are still low velocity objects whose angular motion can be described by the slow particle approximations with small or large $b_Mv$. Therefore a proper application of our results will predict the outgoing angle of HVS, or even better--if these angles are deduced from other means or observed, reveal information such as central GC mass or the impact parameter.

The second circumstance that we can conceive is to use the dependence of the deflection angle on the particle velocity to measure the mass of the particle. One attempt is to measure the masses of neutrino emitted from supernova and hence determine the neutrino mass hierarchy.
In a supernova, the energy spectrum of neutrinos are fixed by the explosion mechanism to be around 10-20 MeV depending on details of the model. For two neutrinos mass eigenstates $|\nu_1\rangle$ and $|\nu_2\rangle$ with the same energy $E_\nu$, the difference in their masses $m_1$ and $m_2$ will lead to a velocity difference. The corresponding difference in the weak lensing deflection angle between a neutrino and the light ray $\theta^\prime_{\pm, \nu_i}-\theta_{\pm}$ and between the two neutrino eigenstates $\theta^\prime_{\pm, \nu_1}-\theta^\prime_{\pm,\nu_2}$, to the leading order of the ${\mathcal O}(m_i^2/E^2)$ can be obtained from Eq. \refer{defangfw} as
\bea
&&\theta^{\prime}_{\pm, \nu_i}-\theta_\pm=\pm\frac{m_i^2}{E^2} \frac{4Mr_i}{\sqrt{r_f(r_i+r_f)[\beta^2r_f(r_i+r_f)+16Mr_i]}},\label{nldiff}\\
&&\theta^\prime_{\pm, \nu_1}-\theta^\prime_{\pm,\nu_2}=\pm\frac{m_1^2-m_2^2}{E^2} \frac{4Mr_i}{\sqrt{r_f(r_i+r_f)[\beta^2r_f(r_i+r_f)+16Mr_i]}}. \label{nndiff}\eea
If one had a neutrino observatory with enough angular resolution, then from Eq. \refer{nldiff} clearly the absolution value of the mass $m_i$ can be determined by comparing with light deflection. It is also seen from Eq. \refer{nndiff} that by inspecting the sign of $\theta^\prime_{\pm, \nu_1}-\theta^\prime_{\pm,\nu_2}$, the neutrino mass hierarchy can be determined.
Moreover if heavier and slower particles, such as sterile neutrinos and WIMPs in some theories, can be detected,  angle differences \refer{nldiff} and \refer{nndiff} can also be considered for them to deduce their properties such as their mass and mass differences.

Finally, our consideration of the trajectory bending and  gravitational lensing is for a Schwarzschild spacetime. Based on previous work on GL of generalized spacetime with a rotation symmetry \cite{Bozza:2002zj}, we expect that the continuity of the trajectories for massless and massive particles, the velocity correction to bending angle and GL can be similarly stuided.

\ack
We appreciate the discussion with Dr. Shun Zhou and Dr. Zonghong Zhu.  The work of X. Liu and J. Jia are supported by the Chinese SRFDP 20130141120079, NNSF China 11504276 \& 11547310, MST China 2014GB109004 and NSF Hubei ZRY2014000988. The work of N. Yang is supported by the NNSF China 31401649 \& 31571797.

\appendix
\section{Convention of elliptic functions}
In this work we used the symbolic analysis software {\it Maple}. Consequently the definitions of the elliptic functions in this work followed its convention:
\bea
&&F(z,k)=\int_0^z \frac{1}{\sqrt{1-t^2}\sqrt{1-k^2t^2}}dt,\\
&&K(k)=F(1,k),\\
&&E(z,k)=\int_0^z \frac{\sqrt{1-k^2t^2}}{\sqrt{1-t^2}}dt.
\eea

%\bibliographystyle{unsrt}
%\bibliography{/home/thiago/bibtex/articles,/home/thiago/bibtex/books}

\end{document}